\documentclass[preprint,12pt]{elsarticle}

\usepackage{graphicx}

\usepackage{amssymb}

\usepackage{amsmath}

\usepackage{upgreek}   

\newcommand{\refeqn}[1]{Eq.~(\ref{#1})}

\journal{Advances in Planar Lipid Membranes \& Liposomes}

\begin{document}

\begin{frontmatter}



  \title{Permeability of phospholipid membrane for small polar
    molecules determined from osmotic swelling of giant phospholipid
    vesicles}


\author[oi]{Primo\v{z} Peterlin}
\address[oi]{Institute of Oncology, Zalo\v{s}ka 2, SI-1000 Ljubljana,
  Slovenia}

\author[ibf]{Vesna Arrigler}
\address[ibf]{Institute of Biophysics, Faculty of Medicine, University
  of Ljubljana, Lipi\v{c}eva 2, SI-1000 Ljubljana, Slovenia}

\author[tau]{Haim Diamant}
\author[tau]{Emir Haleva}

\address[tau]{Raymond and Beverly Sackler School of Chemistry, Tel
  Aviv University, Tel Aviv 69978, Israel}

\begin{abstract}
  A method for determining permeability of phospholipid bilayer based
  on the osmotic swelling of micrometer-sized giant unilamellar
  vesicles (GUVs) is presented as an alternative to the two
  established techniques, dynamic light scattering on liposome
  suspension, and electrical measurements on planar lipid bilayers.
  In the described technique, an individual GUV is transferred using a
  micropipette from a sucrose/glucose solution into an isomolar
  solution containing the solute under investigation.  Throughout the
  experiment, vesicle cross-section is monitored and recorded using a
  digital camera mounted on a phase-contrast microscope.  Using a
  least-squares procedure for circle fitting, vesicle radius $R$ is
  computed from the recorded images of vesicle cross-section.  Two
  methods for determining membrane permeability from the obtained
  $R(t)$ dependence are described: the first one uses the slope of
  $R(t)$ for a spherical GUV, and the second one the $R(t)$ dependence
  around the transition point at which a flaccid vesicle transforms
  into a spherical one.  We demonstrate that both methods give
  consistent estimates for membrane permeability.
\end{abstract}


\end{frontmatter}

\tableofcontents


\section{Introduction}
\label{sec:intro}

The cell membrane physically separates the cell interior from the
environment.  The membrane is selectively permeable: it permits the
passage of some species of molecules or ions to- and from the cell,
while it blocks the transport for other species.  Around 1900, while
studying the working of general anaesthesia, Hans Meyer in Marburg
\cite{Meyer:1899} and Ernest Overton in Z\"urich \cite{Overton:1901}
independently devised a simple rule to predict membrane permeability.
They established that the more a molecule species is soluble in lipid,
the greater the cell permeability for this molecule species is.  While
this rule cannot account for transport processes not known at that
time, such as those mediated by membrane carriers, channels
(\emph{e.g.}, aquaporins \cite{Preston:1992}) or pumps, nor does it
acknowledge membrane inhomogeneities, such as rafts \cite{Pike:2006}
present in the biological membrane, the Meyer-Overton rule seems to
withstand the test of time \cite{AlAwqati:1999,Missner:2009}.  At the
same time, there is a continuing interest in both basic physics of
general anaesthesia \cite{Heimburg:2007} and passive membrane
permeability \cite{Nagle:2008}.


While a significant corpus of publications on permeability properties
of biological cells has been accumulated since Meyer and Overton,
experimental access to an isolated lipid bilayer only became available
in the early 1960s, when a technique for preparation of thin films
separating two aqueous compartments has been devised, known as the
black lipid membrane \cite{Mueller:1962}.  An arguably even more
influential technique which was developed at roughly the same time
allowed for producing artificial lipid vesicles, or liposomes
\cite{Bangham:1968}.  Vesicles are osmotically sensitive structures
which swell and shrink in response to changed osmotic conditions.
Even though the sub-micrometer liposomes which are the easiest to
produce cannot be visualized directly, their size can be estimated
via light scattering \cite{Cohen:1972} (see \cite{Verkman:1995} for a
review on the optical methods in determining membrane permeability).

Both model lipid bilayer systems -- planar lipid bilayers and
liposomes -- have proved extremely fruitful in the studies of membrane
permeability \cite{Tien:1974}.  Planar lipid bilayers are well suited
to electrical characterization, as the two chambers separated by the
bilayer are both easily accessible, which allows simple placement of
macroscopic electrodes.  Finkelstein \cite{Finkelstein:1976} measured
the permeability of a planar lipid bilayer for water and seven other
non-electrolytes in an attempt to resolve the mechanism by which
neutral molecules and ions permeate the membrane. One possibility is
the solubility-diffusion mechanism, which assumes that the permeating
species dissolves in the hydrophobic membrane, diffuses across, and
leaves by redissolving into the other aqueous compartment.  Another
possibility is that permeation occurs through hydrated transient
defects, which appear as a result of thermal fluctuations.  That study
concluded that both water and non-electrolytes cross the membrane
through the solubility-diffusion mechanism.  A similar but more
comprehensive study was repeated a few years later by Orbach and
Finkelstein \cite{Orbach:1980}.  Walter and Gutknecht
\cite{Walter:1986} examined the correlation between the membrane
permeability for 22 solutes and their partition coefficient between
water and any of the four examined organic solvents, and found a very
high correlation with hexadecane and olive oil, and a less pronounced
correlation in the case of octanol and ether.

In an early work \cite{Bangham:1967}, Bangham and coworkers made use
of the fact that the total volume of liposomes in the suspension is
proportional to the reciprocal of the optical extinction, which
allowed them to examine the permeability of membrane for water and
various solutes via osmotic swelling and shrinking of liposomes.  As
early as 1933 Jacobs noticed \cite{Jacobs:1933} that the volume of a
cell transferred into a solution of permeant solute transiently
decreases, reaches some minimal value, and then starts increasing.
Sha'afi and coworkers \cite{Shaafi:1970} employed this phenomenon and
the Kedem-Katchalsky formalism \cite{Kedem:1958} to compute the
permeability of erythrocyte membrane for urea.  De Gier and coworkers
used the initial slope of the reciprocal of the optical extinction for
determining membrane permeability of liposomes \cite{DeGier:1971b}.
Hill and Cohen \cite{Hill:1972} brought the ``minimal volume''
technique to the experiment with liposomes as well.  A comprehensive
review of the use of liposomes in membrane permeability studies is
given by de Gier \cite{Degier:1993}.  Using established techniques,
Paula \emph{et al.} \cite{Paula:1996} did an extensive study in
another attempt to resolve the standing debate between the
solubility-diffusion mechanism and the hydrated transient defects as a
primary pathway, and concluded that except for ion permeability of
lipid bilayers composed of phospholipids with short chain lengths,
solubility-diffusion mechanism seems to be the dominant effect.
Examining the known phenomenon that upon transfer into a hypotonic
medium, vesicles swell, and, if the gradient is large enough, burst
and expel part of their content, Shoemaker and Vanderlick studied the
influence of membrane composition on the extent of leakage
\cite{Shoemaker:2002}, and found out a correlation between the
membrane resistance to burst and its stretching modulus.

Both planar lipid bilayers and liposomes as model bilayer systems have
their drawbacks, too.  The original ``brush'' technique of producing
planar lipid bilayers has been limited by pockets of residual solvent
trapped between the two bilayer leaflets, which affects membrane
properties.  While an improved deposition method \cite{Montal:1972}
virtually eliminated this problem, the limited lifetime of the
membranes-- most often less than one hour -- remains a persisting
problem which limits the duration of the experiment.  Liposomes, on
the other hand -- in particular large unilamellar vesicles (LUVs) with
a diameter of 100--200~nm -- have proved to be extremely stable.
The interpretation of dynamic light scattering (DLS) experiments of
osmotic shrinking of LUVs is not trivial, as it involves the
transformation of shapes predicted by the area difference elasticity
model \cite{Seifert:1997} to the hydrodynamic radius, which is
characterized by DLS \cite{Pencer:2001,Pencer:2003}.  Also, it has
been argued that LUVs, which consist exclusively of high-curvature
membrane regions, serve as a poor model of biological cell membranes.
An attempt to resolve the possible dependence of permeability on the
membrane curvature \cite{Brunner:1980} was inconclusive, as the
authors ascribe the observed differences in permeability to the
problems they experienced with planar lipid membranes.  Finally, an
effect which occurs in both systems, but is more prominent with the
planar lipid bilayer, is the unstirred layer effect \cite{Barry:1984}.
In general, the concentration of solute adjacent to the membrane
differs from its concentration in bulk.  It is the concentration of
solute immediately adjacent to the membrane which determines the
permeation of solute across the membrane, while the concentration in
bulk is the one that is usually known.  In both the experiment design
and the interpretation of the experimental findings, one needs to be
aware of this discrepancy.

While both planar lipid bilayers and liposomes have been used as model
membrane systems for studies of membrane permeability since 1960s,
studies employing GUVs, which allow for a direct visualization of the
process, appeared decades later \cite{Boroske:1981}, chiefly due to a
lack of suitable techniques for preparation and manipulation of GUVs
in those early days.  In the paper by Boroske \emph{et al.}
\cite{Boroske:1981}, the authors describe the experiment in which GUVs
were prepared in pure water and subsequently transferred into a
solution of either glucose or NaCl (concentrations used ranged from
1.5--20~mM) while their size was monitored using phase-contrast
microscopy.  Upon transfer, vesicles shrank in size; the process of
shrinking depended on the vesicle size.  Vesicles with radius $R
\lesssim 10\; \upmu\textrm{m}$ shrank with a linear time dependence:
$R(t) = R_0 - \bar{V}_w P \Delta c\, t$, where $\bar{V}_w$ is the
molar volume of water, $P$ is membrane permeability for water (water
filtration coefficient), $\Delta c$ the solute concentration
difference, and $t$ time. From measured data, the authors inferred the
water filtration coefficient, $P =
41\;\upmu\textrm{m}/\textrm{s}$. Larger vesicles ($r \gtrsim 10\;
\upmu\textrm{m}$) underwent a phase of ``instability'' in which the
vesicle was flaccid, and after which a spherical shape was reestablished.
The authors dismissed the idea of dissolving lipid molecules into the
outer medium as a plausible explanation for the apparent decrease of
vesicle surface area, in particular since they also noticed formation
of smaller satellite spherical vesicles, seemingly connected to the
mother vesicle.  Instead, they proposed a mechanism of concerted
flipping of lipid molecules from the inner membrane leaflet into the
outer membrane leaflet, induced by the flow of water.  Recently,
membrane permeability has also been studied of GUVs made of block
copolymers \cite{Wang:2010,Carlsen:2011}.

The remaining of this chapter is structured as follows. First, the
experimental section introduces the system and presents the immediate
experimental results.  A section on the theory of membrane
permeability offers a review of the few selected phenomenological
models for membrane permeability, with a special emphasis on the
influence of membrane elasticity and the swelling-burst cycle. The
section concludes with a theory of the continous transition between
the ``ironing'' and the stretching regimes of the osmotic swelling of
a vesicle.  The section on experimental analysis demonstrates the
calculation of membrane permeability based on the swelling-burst
cycle, and compares its result with the calculation based on the
analysis of the transition between the ``ironing'' and the stretching
regimes.  We conclude with a discussion of the merits and limitations
of the presented method.

\section{Experimental section}
\label{sec:experiment}

\subsection{Materials and methods}

D-(+)-glucose, D-(+)-sucrose, glycerol, urea, and ethylene glycol were
purchased from Fluka (Buchs, Switzerland).  Methanol and chloroform
were purchased from Kemika (Zagreb, Croatia).
1-palmitoyl-2-oleoyl-\textit{sn}-glycero-3-phosphocholine (POPC) was
purchased from Avanti Polar Lipids (Alabaster, USA).  All the
solutions were prepared in double-distilled sterile water.


A suspension of POPC GUVs in 0.1 or 0.2~mol/L 1:1 sucrose/glucose
solution was prepared using an electroformation method, described in
Ref.~\citenum{Angelova:1986} with some modifications
\cite{Heinrich:1996,Peterlin:2008a}.  Lipids were dissolved in a
mixture of chloroform/methanol (2:1, v/v) to a concentration of
1~mg/mL.  A volume of 25~$\upmu$L of the lipid solution was spread
onto a pair of Pt electrodes and dried under reduced pressure (water
aspirator; $\approx 60$~mmHg) for 2~hours.  The electrodes were then
placed into an electroformation chamber, which was filled with 0.1 or
0.2~mol/L sucrose.  AC current (8~V, 10~Hz) was applied, and the
voltage and frequency were reduced in steps to the final values of 1~V
and 1~Hz \cite{Peterlin:2008a}.  Subsequently, the chamber was first
drained into a beaker and then flushed with an equal volume of
isomolar glucose solution, thus resulting in a suspension of GUVs
containing entrapped sucrose in a 1:1 sucrose/glucose solution, which
increases the contrast in a phase contrast setup and facilitates
vesicle manipulation \cite{Dimova:2006}.  This procedure yields mostly
spherical unilamellar vesicles, with diameters of up to 100~$\upmu$m.


An inverted optical microscope (Nikon Diaphot 200, objective 20/0.40
Ph2 DL) with micro-manipulating equipment (Narishige MMN-1/MMO-202)
and a cooled CCD camera (Hamamatsu ORCA-ER; C4742-95-12ERG), connected
{\it via} an IEEE-1394 interface to a PC running Hamamatsu Wasabi
software, was used to obtain phase contrast micrographs.  In the
streaming mode, the camera provides $1344\times 1024$ 12-bit grayscale
images at a rate of 8.9~images/s.

In the experiment, an individual spherical GUV is selected, fully
aspirated into a glass micropipette whose inner diameter exceeds the
vesicle's diameter, and transferred from a solution containing solutes
of very low membrane permeability (1:1 glucose/sucrose) into an
iso-osmolar solution of a more permeable solute (glycerol, urea, or
ethylene glycol), where the content of the micropipette is released,
and the micropipette is subsequently removed.  Vesicle response is
recorded using a CCD camera mounted on the microscope.

\subsection{Experimental results}

\begin{figure}[t]
  \centering\includegraphics[width=0.9\linewidth]{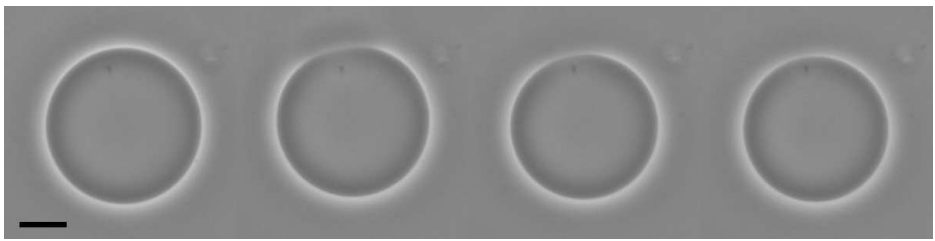}
  \caption{A sequence of micrographs showing an individual vesicle
    burst.  Vesicle radius before the burst was 32.69~$\upmu$m;
    immediately after the burst, it shrunk to 30.47~$\upmu$m.  The
    images were taken at 0.11~s intervals.  The bar represents
    20~$\upmu$m.}
  \label{fig:burst}
\end{figure}

Upon transfer into a solution of permeating solute, vesicles start to
swell until the membrane critical strain is reached.  At that point,
the membrane ruptures, and the vesicle ejects part of its internal
solution (figure~\ref{fig:burst}).  After the burst, the membrane
reseals, and another cycle of swelling starts.  The observed sequences
of swelling-burst cycles ranged from 3 to over 40 successive bursts,
with an average in our sample being 15.4.

\begin{figure}[t]
  \centering\includegraphics[width=0.9\linewidth]{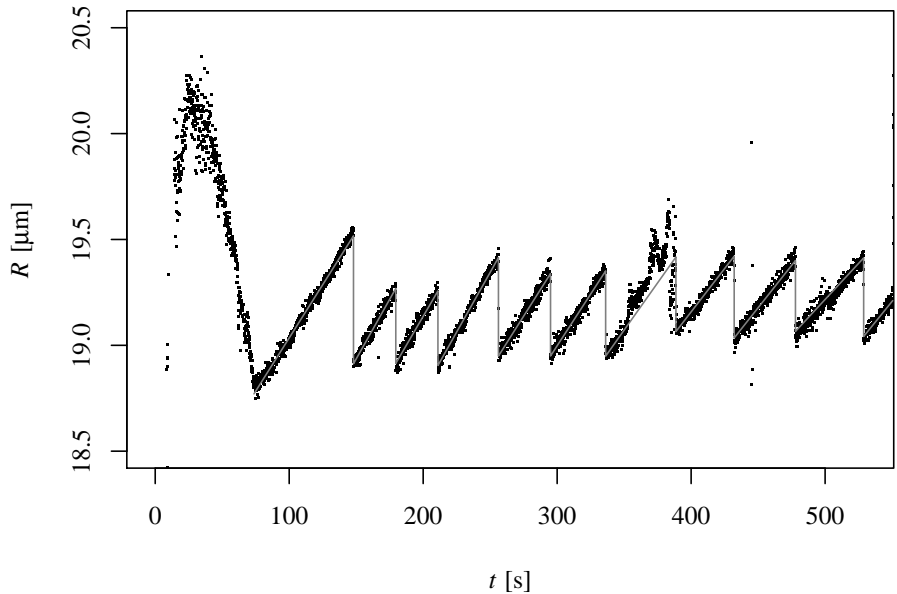}
  \caption{Time dependence of the radius of a vesicle cross-section
    upon transfer from a 0.1~mol/L 1:1 glucose/sucrose solution into a
    0.1~mol/L solution of ethylene glycol.  Clearly visible are the
    initial maximum, followed by a saw-tooth pattern of repeated
    swelling-burst cycles.}
  \label{fig:burst-seq}
\end{figure}

For a quantitative analysis, the radius of the vesicle cross-section
in each recorded micrograph was determined with a GNU Octave script
using a least-squares procedure for circle fitting
\cite{Peterlin:2009a}.  Figure~\ref{fig:burst-seq} shows a typical
time course of vesicle radius after transfer.  In about 80\% of all
cases, we observed a transient maximum of radius shortly after the
transfer.  We attribute this phenomenon to a slight hypertonicity of
the target solution.  This causes the vesicle to deflate rapidly, the
rate of volume change being determined by the vesicle size and the
permeability of phospholipid membrane for water.  A deflated vesicle
changes its shape from spherical into a shape which can be
approximated with an oblate spheroid.  Small deviations from this
shape, which originate from the effects of gravity
\cite{Dobereiner:1997}, are neglected here.  Due to gravity, the axis
of rotational symmetry is aligned with the vertical, which, in our
experimental setup, also coincides with the direction of the optical
axis.  Thus, the observed cross-section radius is the oblate spheroid
longer semiaxis, and consequently it increases with the decreasing
volume, while the membrane area remains unchanged.

Concurrently with the initial vesicle deflation due to the slight
hypertonicity, diffusion of permeating solute into the vesicle
interior, accompanied by the osmotic influx of water, also takes
place.  The rate of this latter process is however dominated by the
permeability of the membrane for the given solute (in our case,
glycerol, urea, or ethylene glycol), and is thus much slower than the
initial rapid deflation due to the mismatch in tonicity.  With time,
however, the initial trend of vesicle deflation is overturned and the
vesicle starts inflating again.  Inflation can be qualitatively
divided into two phases.  In the first phase, the vesicle is still
partially deflated, and the influx of solute and the accompanying
osmotic influx of water increases the vesicle volume, all while the
membrane area remains unchanged.  This ``ironing'' phase is
characterized by the decrease of the radius of vesicle cross-section.
At some point, the vesicle reaches a spherical shape, which can be
observed as a local minimum of the cross-section radius.

Further permeating of solute into the vesicle interior, accompanied by
the osmotic influx of water, causes the vesicle to inflate while
maintaining a spherical shape.  In this phase, the increase of vesicle
volume can be observed as an increase of vesicle cross-section.  In
order to accomodate the increased vesicle volume, the membrane needs
to stretch.  This stretching process continues until the critical
strain for the membrane is reached.  At this point, the membrane
ruptures, and part of the vesicle interior is ejected outside.  When
the vesicle volume is thus reduced, the membrane reseals again.  As
the concentration difference for the permeating solute persists, which
serves as a driving force for the diffusion of the permeating solute
into the vesicle interior, this means that at the same moment, the
vesicle volume starts increasing again, and another swelling cycle
commences.  Repeated swelling-burst cycles give yield to the
characteristic saw-tooth pattern, when the vesicle radius is plotted
against the elapsed time (figure~\ref{fig:burst-seq}).

\begin{figure}[t]
  \centering\includegraphics[width=0.7\linewidth]{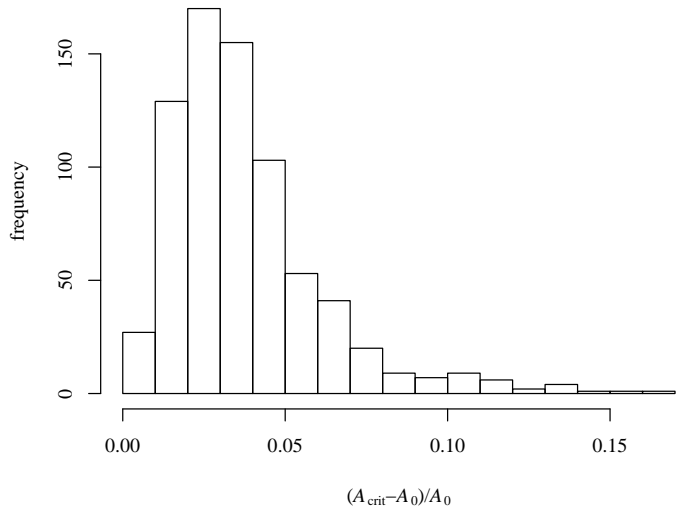}
  \caption{A histogram of experimentally determined critical strain,
    defined as $(R_\mathrm{crit}/R_0)^2 - 1$, where $R_0$ is the
    initial and $R_\mathrm{crit}$ the final, or critical, radius for
    each of the swelling cycles.}
  \label{fig:critical-strain}
\end{figure}

In total, 47 recordings of vesicle transfer from a 1:1 sucrose/glucose
solution into an isomolar solution of glycerol, urea, or ethylene
glycol were selected for further analysis.  Out of these, 15 transfers
were into glycerol (5 at 0.1~M and 10 at 0.2~M), 15 transfers into
urea (8 at 0.1~M and 7 at 0.2~M), and 17 transfers into ethylene
glycol (12 at 0.1~M and 5 at 0.2~M).  Figure~\ref{fig:critical-strain}
shows the critical strain $(A_\mathrm{crit}-A_0)/A_0 =
(R_\mathrm{crit}/R_0)^2 - 1$, where $R_0$ is the radius of a relaxed
spherical vesicle, and $R_\mathrm{crit}$ is the radius of a critically
strained vesicle.  The median critical strain obtained for a total of
738 recorded vesicle bursts is 0.033, and mean critical strain is
0.038 with standard deviation 0.024.

A published value for the critical strain is around 0.04
\cite{Evans:1990,Bloom:1991}.  Our own estimate based on the first
burst in the sequence alone is a little higher: $0.055\pm0.02$
\cite{Peterlin:2012}.  This may indicate that the membrane might not
always perfectly reseal, and that a local defect present in the
membrane makes it more likely to rupture at a lower strain.

\section{Theoretical section}
\label{sec:theory}

In this section, we will briefly review a few theoretical models used
for the description of solute permeation across cell membrane,
starting from the most general Kedem-Katchalsky formalism, based on
nonequilibrium thermodynamics \cite{Katchalsky:1965}, then showing how
in the case of separate pathways for water and solute transport it can
be simplified into a two-parameter model, and subsequently when a
two-parameter model can be simplified into a one-parameter model in
the case when the membrane permeability for water greatly exceeds its
permeability for solute.  We will discuss the influence of finite
membrane elasticity on the apparent membrane permeability, review the
work on the repetitive swelling-burst cycles, and conclude with the
theory of the continuous transition between the ``ironing'' and the
stretching regime of a vesicle which swells osmotically due to the
permeation of a permeable solute.

\subsection{Kedem-Katchalsky formalism}

Starting from the principles of non-equilibrium thermodynamics, Kedem
and Katchalsky arrived at a model with three parameters
\cite{Kedem:1958}.  A simple qualitative argument can be offered why
no less than three parameters are required for the description of the
transport of solvent and solute across a membrane.  In the simpler
case of free diffusion of a single solute in a solvent, molecules of
solute move relative to the molecules of solvent, and a single
parameter -- diffusion coefficient -- is required to describe the
friction between the solvent and the solute.  In the case of membrane
transport, two additional coefficients are required, one describing
the friction between the molecules of solvent and the membrane, and
another describing the friction between the molecules of solute and
the membrane.

As their starting point, Kedem and Katchalsky \cite{Kedem:1958} choose
the rate of entropy production during the permeation of solute and
water across the membrane separating the interior compartment from the
exterior,
\begin{equation}
  \label{eq:entropy-rate}
  \frac{\mathrm{d}S}{\mathrm{d}t} = 
  \frac{1}{T} \left(\mu_w^e - \mu_w^i\right) 
  \frac{\mathrm{d}N_w^i}{\mathrm{d}t} +
  \frac{1}{T} \left(\mu_s^e - \mu_s^i\right) 
  \frac{\mathrm{d}N_s^i}{\mathrm{d}t} \; .
\end{equation}
Here, $\mu_s$ and $\mu_w$ denote the chemical potential of solute and
solvent (water), respectively, and $\mathrm{d}N_s^i/\mathrm{d}t$ and
$\mathrm{d}N_w^i/\mathrm{d}t$ denote the number of moles of solvent
and water entering the interior compartment per unit time.

Formulating \refeqn{eq:entropy-rate} in terms of a dissipation function
per unit area, $\Phi = (T/A)\, \mathrm{d}S/\mathrm{d}t$, one obtains
\begin{equation}
  \label{eq:dissipation-nsnw}
  \Phi = \left(\mu_w^e - \mu_w^i\right) \dot{n}_w +
  \left(\mu_s^e - \mu_s^i\right) \dot{n}_s \; .
\end{equation}
Here, we have denoted $\dot{n}_w = (1/A)\,
\mathrm{d}N_w^i/\mathrm{d}t$ for water and accordingly for the solute.
\refeqn{eq:dissipation-nsnw} is a particular case of dissipation
function, which in general assumes the form $\Phi = \sum_i J_i X_i$,
where $J_i$ represent flows and $X_i$ represents the corresponding
conjugate ``forces''. In \refeqn{eq:dissipation-nsnw}, the differences
in chemical potential act as conjugate forces.  We shall now rewrite
\refeqn{eq:dissipation-nsnw} in a way where the driving forces will be
quantities easier to evaluate experimentally.

For an ideal solution, the difference in chemical potential can be
expressed as
\begin{equation}
  \label{eq:ideal-solution}
  \mu^e - \mu^i = \bar{V}\,\Delta p + \alpha\, \Delta \ln x \; ,
\end{equation}
where $\bar{V}$ denotes the partial molar volume of the constituent,
$x$ its mole fraction, $\Delta p$ the pressure difference between the
external and the internal compartment, and $\alpha = RT$, with $R$
being the molar gas constant and $T$ the absolute temperature.
Assuming a dilute solution, where the volume fraction $\varphi$ of the
solute is low, $\varphi = c_s \bar{V}_s \ll 1$, with $c_s = N_s/V_w$
being the molal concentration of solute, one can rewrite
\refeqn{eq:ideal-solution} for the solute:
\begin{equation}
  \label{eq:ideal-solution-solute}
  \mu_s^e - \mu_s^i = \bar{V}_s\,\Delta p + \alpha\frac{\Delta c_s}{c_s} \; ,
\end{equation}
where $\Delta c_s = c_s^e - c_s^i$ is the difference and $c_s = (c_s^e
+ c_s^i)/2$ is the average of the concentration of solute in both
compartments.  An analogous relation can be written for the solvent,
\begin{equation}
  \label{eq:ideal-solution-solvent}
  \mu_w^e - \mu_w^i = \bar{V}_w\,\Delta p - \alpha\frac{\Delta c_s}{c_w} \; ,
\end{equation}
with $c_w = (1-\varphi)/\bar{V}_w \approx 1/\bar{V}_w$.  Introducing
\refeqn{eq:ideal-solution-solute} and
\refeqn{eq:ideal-solution-solvent} into \refeqn{eq:dissipation-nsnw}
and rearranging, one arrives at another expression for the dissipation
function:
\begin{equation}
  \label{eq:dissipation-func}
  \Phi = \left(\dot{n}_w \bar{V}_w + \dot{n}_s \bar{V}_s \right) \Delta p +
  \left(\frac{\dot{n}_s}{c_s} - \frac{\dot{n}_s}{c_s}\right) \alpha\,\Delta c_s
  \; .
\end{equation}
In \refeqn{eq:dissipation-func}, $\Phi$ is expressed in terms of forces
commonly used in the permeability studies: $X_v = \Delta p$ is the
hydrostatic pressure, and $X_D = \alpha\,\Delta c_s$ is the osmotic
pressure.  The conjugate flows are $J_v = \dot{n}_w \bar{V}_w +
\dot{n}_s \bar{V}_s$, the total volume flow per unit area, and $J_D =
\dot{n}_s/c_s - \dot{n}_s/c_s$, the relative velocity of solute with
respect to solvent, which serves as a measure of exchange flow.

It is assumed that each flow $J$ present in the system in general
depends on all the forces $X$ acting in the system, and if the forces
are sufficiently small, the relationship is linear:
\begin{align*}
  J_1 & = L_{11} X_1 + L_{12} X_2 \\
  J_2 & = L_{21} X_1 + L_{22} X_2
\end{align*}
Here, $L_{ik}$ are phenomenological transport coefficients.  The
Onsager reciprocity relations \cite{Onsager:1931a,Onsager:1931b} state
that the matrix of transport coefficients is a diagonal one, $L_{ki} =
L_{ik}$.

Applying this formalism to our case, we obtain
\begin{alignat}{3}
  J_v &= L_p   &\Delta p &+{}& L_{pD} &\,\alpha\,\Delta c_s \; , \\
  J_D &= L_{pD} &\Delta p &+{}& L_{D}  &\,\alpha\,\Delta c_s  \; ,
\end{alignat}
where we have already taken into account $L_{Dp} = L_{pD}$. The second
law of thermodynamics requires that the diagonal terms are
non-negative, while the off-diagonal terms are only constrained by the
relation $L_p L_D - L_{pD}^2 > 0$.  It is the off-diagonal terms
though which are responsible for a concentration difference producing
a volume flow, or vice versa.

For easier comparison with the experimental results, it is convenient
to transform $\{L_p, L_D, L_{pD}\}$ to another set of
coefficients. One of them is the reflection coefficient $\sigma$,
introduced with
\begin{equation}
  \label{eq:reflection-coef}
  L_{pD} = - \sigma L_p \; .
\end{equation}
Two special cases which can be considered include a non-selective
membrane ($\sigma = 0$) and an ideally selective membrane, permeable
only for the solute ($\sigma = 1$).  Another parameter commonly
defined is the mobility of the solute $\omega$:
\begin{equation}
  \label{eq:mobility-solute}
  \omega = \frac{L_p L_D - L_{pD}^2}{L_p} c_s = (L_D - L_p \sigma^2) c_s \; .
\end{equation}
It can be shown that $\omega$ is chosen in such a way that for
permeability measurement at constant volume ($J_v = 0$), one can write
$\dot{n}_s = \omega R T\, \Delta c_s$.

Expressing the volume flow and the exchange flow in terms of $\{L_p,
\omega, \sigma\}$, one obtains
\begin{align}
  J_v &= L_p \Delta p - \sigma L_p R T\, \Delta c_s \; , 
  \label{eq:jv-binary} \\
  \dot{n}_s &= (1 - \sigma) L_p c_s \Delta p + \left[ \omega - 
    \sigma (1 - \sigma) L_p c_s \right] R T\, \Delta c_s \; .
  \label{eq:ns-binary}
\end{align}

\refeqn{eq:jv-binary} and \refeqn{eq:ns-binary} obtained above pertain
to a two-component system, in which binary solutions of the same
solvent and solute are separated by a membrane. In order to allow for
a comparison with the experimental results, we need to consider a
slightly more complicated system, comprising of a membrane separating
two ternary solutions of the same solvent and two different solutes.
The membrane is permeable for the solvent and one of the solutes, but
impermeable for the other solute. As in the previous example, we treat
the solution as ideal.

For a dilute solution, the difference in the chemical potential for
the solvent (\refeqn{eq:ideal-solution-solvent}) can now be written as
\begin{equation}
  \label{eq:ternary-solvent}
  \Delta \mu_w = - \alpha\frac{\Delta c_s}{c_w} - \alpha\frac{\Delta c_n}{c_w} 
  + \bar{V}_w\,\Delta p \; ,
\end{equation}
where the index $s$ pertains to the permeating solute, and the index
$n$ to the non-permeating solute.  While
\refeqn{eq:ideal-solution-solute} remains valid, we have to consider
the contribution of the non-permeating solute to the difference in
osmotic pressure,
\begin{equation}
  \label{eq:osmotic-nonperm}
  \Delta\Pi_n = \alpha\, \Delta c_n \; .
\end{equation}

Considering the same flows as before, one can write the expression for
the dissipation function \refeqn{eq:dissipation-nsnw}, and when
substituting into it the expressions for the difference in the
chemical potential for the solute, \refeqn{eq:ideal-solution-solute}, and
solvent, \refeqn{eq:ternary-solvent}, one obtains the expressions for
the forces conjugate to the flows:
\begin{align}
  X_v &= \Delta p - \Delta\Pi_n \; , \label{eq:ternary-Xv} \\
  X_D &= \alpha\, \Delta c_s + \varphi\,\Delta\Pi_n \; . \label{eq:ternary-XD}
\end{align}
Using the expressions (\ref{eq:ternary-Xv},\ref{eq:ternary-XD}), one
can obtain expressions analogous to
(\ref{eq:jv-binary},\ref{eq:ns-binary}), obtained in the case of a
binary solution:
\begin{align}
  J_v &= L_p (\Delta p - \Delta\Pi_n) - 
  \sigma L_p (\alpha\, \Delta c_s + \varphi\,\Delta\Pi_n) \; , 
  \label{eq:jv-ternary} \\
  \dot{n}_s &= (1 - \sigma) L_p c_s (\Delta p - \Delta\Pi_n) + 
  \left[ \omega - \sigma (1 - \sigma) L_p c_s \right] 
  (\alpha\, \Delta c_s + \varphi\,\Delta\Pi_n) \; .
  \label{eq:ns-ternary}
\end{align}
A more condensed expression of \refeqn{eq:ns-ternary} can be obtained
if \refeqn{eq:jv-ternary} is taken into account:
\begin{equation}
  \label{eq:ns-ternary-var}
  \dot{n}_s = (1 - \sigma) c_s J_v + 
  \omega (\alpha\, \Delta c_s + \varphi\,\Delta\Pi_n) \; .
\end{equation}
In dilute solutions, $\varphi\,\Delta\Pi_n$ is often negligible with
respect to $\alpha\,\Delta c_s$.  On the other hand, the contribution
$\Delta\Pi_n$ is important.  In many biologically relevant
experiments, $\Delta p = 0$ while $\Delta\Pi_n \neq 0$.

\subsection{Two-parameter model}

Through the reflection coefficient $\sigma$, the Kedem-Katchalsky
formalism resolves the competition between solvent and solute being
transported through a shared pathway, e.g., a cotransporting channel
permeable to both the solute and the solvent.  The formalism itself,
however, applies to any simple transport problem, regardless of
whether a cotransporting channel is present or not
\cite{Kleinhans:1998}.  In the latter case, not all three parameters
$\{L_p, \omega, \sigma\}$ are independent.  It can be shown
\cite{Kedem:1958} that in this case, $\sigma$ can be written as:
\begin{equation}
  \label{eq:sigma-2p}
  \sigma = 1 - \frac{\omega\bar{V}_s}{L_p} \; .
\end{equation}
Introducing \refeqn{eq:sigma-2p} into
Eqs.~(\ref{eq:jv-ternary},\ref{eq:ns-ternary-var}), one obtains the
transport equations for the case where the solute and the solvent do
not compete for the same cotransporting channel, e.g., in the case
where they both diffuse through the phospholipid bilayer.  Along the
way, we will use the following simplifications: $\alpha\,\Delta c_s \gg
\varphi\,\Delta\Pi_n \approx 0$, $\Delta p = 0$, and introduce the
notation more appropriate for describing the experimental setup: $J_v
= (1/A)\,\mathrm{d}V/\mathrm{d}t$, $\dot{n}_s = (1/A)\,
\mathrm{d}N_s/\mathrm{d}t$, $P_s = \omega \alpha$.  Here, $V$ is the total
volume of the internal compartment, $A$ is the area of the membrane,
$N_s$ is the number of moles of permeating solute inside the internal
compartment, and $P_s$ is the permeability of the membrane for the
permeating solute.  Using the described simplifications, the transport
equations (\ref{eq:jv-ternary},\ref{eq:ns-ternary}) can be written
as
\begin{align}
  \frac{\mathrm{d}V}{\mathrm{d}t} &= -L_p \alpha A \left[ (c_n^e + c_s^e) -
    (c_n^i + c_s^i)\right] + P_s \bar{V}_s A (c_s^e - c_s^i) 
  \label{eq:dV-2p} \; , \\
  \frac{\mathrm{d}N_s}{\mathrm{d}t} &= - P_s \bar{V}_s c_s A 
  (c_n^e - c_n^i) + P_s A (c_s^e - c_s^i) \; .
  \label{eq:dNs-2p}
\end{align}
Two terms contribute to the volume change in \refeqn{eq:dV-2p}: the
first one corresponds to the transport of solvent due to the osmotic
pressure gradient, and the second one to the transport of solute.  For
dilute solutions, the second term is much smaller.  Similarly, in
\refeqn{eq:dNs-2p}, the first term is proportional to $\bar{V}_s c_s$,
which makes this term negligible for dilute solutions.

As an example, we will apply the two-parameter model to the case which
corresponds to the experimental setup: a single vesicle filled with a
non-permeating solute is transferred to a reservoir filled with an
isotonic solution of a permeating solute.  Even though the solutions
are isotonic, permeating solute diffuses into the vesicle, thus
causing an osmotic pressure, which is in turn balanced by the inflow
if water.  Two cases can be distinguished: a flaccid vesicle changes
its shape and becomes ever more spherical, while a spherical vesicle
has already reached its limiting shape and can only grow by stretching
the membrane.  In this case, $c_n^e = 0$, $c_s^e = \text{const.}$, and
initially, $c_s^i = 0$, $c_n^i = c_{n0}^i$; which is equal to $c_s^e$.

In the case of a flaccid vesicle, membrane area $A$ is constant, and
the system can be characterized with the vesicle volume $V$ and the
amount of permeating solute in the vesicle interior, $N_s$. Using the
simplifications $V_w \approx V$, which is appropriate for dilute
solutions, \refeqn{eq:dV-2p} and \refeqn{eq:dNs-2p} transform into:
\begin{align}
  \frac{\mathrm{d}V}{\mathrm{d}t} &= -L_p \alpha A \left( c_s^e -
    \frac{N_s}{V} - c_{n0}^i \frac{V_\mathrm{ini}}{V} \right)
  \label{eq:dV-2p-flaccid} \; , \\
  \frac{\mathrm{d}N_s}{\mathrm{d}t} &=  
  P_s A \left(c_s^e - \frac{N_s}{V}\right) \; .
  \label{eq:dNs-2p-flaccid}
\end{align}
Here, $V_\mathrm{ini}$ denotes the initial volume of the vesicle, and
$c_{n0}^i$ denotes the initial partial concentration of the
non-permeating solute inside the vesicle. Often, the permeability of
membrane for water is expressed as water filtration coefficient $P_f$
instead of hydraulic conductivity $L_p$, the two quantities being
bound by the relation $P_f = \alpha L_p/\bar{V}_w$.  Alternatively, one
can use the reduced volume $v = V/V_0$, instead of $V$, with $V_0 =
A^{3/2}/(6\sqrt{\pi})$ being the volume of a sphere with an area equal
to $A$:
\begin{align}
  \frac{\mathrm{d}v}{\mathrm{d}t} &= -\frac{P_f \bar{V}_w A}{V_0} 
  \left( c_s^e - \frac{N_s}{v V_0} - c_{n0}^i \frac{v_0}{v} \right)
  \label{eq:dV-2p-flaccid-v} \; , \\
  \frac{\mathrm{d}N_s}{\mathrm{d}t} &=  
  P_s A \left(c_s^e - \frac{N_s}{v V_0}\right) \; .
  \label{eq:dNs-2p-flaccid-v}
\end{align}
Here, $v_0 = V_\mathrm{ini}/V_0$ is the initial reduced volume of the
vesicle.

The other case is a spherical vesicle. In this case, neither its area
$A$ nor its volume $V$ are constant; both can be, however, expressed
in terms of the vesicle radius $R$, which is a convenient parameter in
this case.  Substituting $A = 4\pi R^2$ and $V = 4\pi R^3/3$ into
\refeqn{eq:dV-2p-flaccid} and \refeqn{eq:dNs-2p-flaccid}, one obtains:
\begin{align}
  \frac{\mathrm{d}R}{\mathrm{d}t} &= -P_f \bar{V}_w 
  \left[ c_s^e - c_s^i - c_{n0}^i \left(\frac{R_0}{R}\right)^3 \right]
  \label{eq:dV-2p-sph} \; , \\
  \frac{\mathrm{d}c_s^i}{\mathrm{d}t} &=  
  \frac{3 P_s}{R} \left(c_s^e - c_s^i\right) \; .
  \label{eq:dNs-2p-sph}
\end{align}
Here, $R_0$ is the initial vesicle radius; $V_0 = 4\pi R_0^3/3$.

\begin{figure}[t]
  \centering\includegraphics{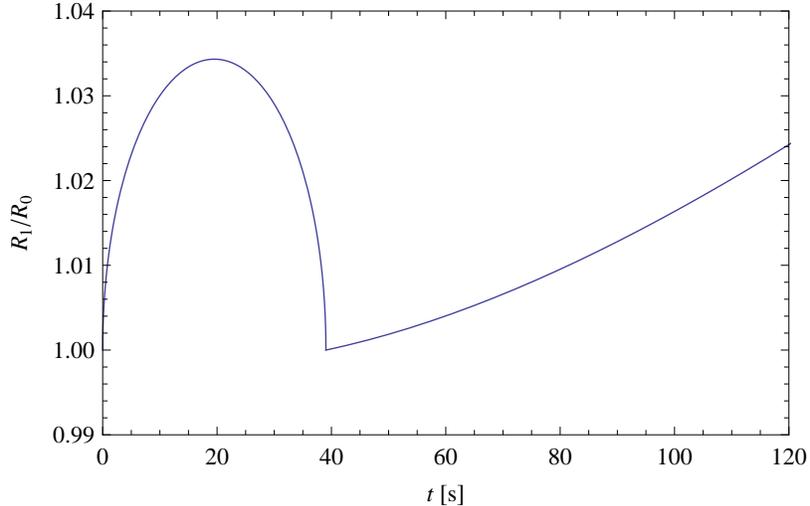}
  \caption{The radius of a vesicle cross-section upon transfer into a
    hypertonic solution of a solute which can permeate through the
    vesicle membrane, as calculated by the two-parameter model.
    Calculation parameters are given in the text.}
  \label{fig:bulge-2p}
\end{figure}

As figure~\ref{fig:burst-seq} shows, both flaccid and spherical
regimes can appear in the course of a single vesicle transfer. The
two-parameter model allows us to mimic the same behaviour.
Figure~\ref{fig:bulge-2p} shows a time course of the radius of a
vesicle cross-section upon transfer into a slightly hypertonic
solution of a solute which can permeate the vesicle membrane. A
spherical vesicle with $R_0 = 10\;\upmu\mathrm{m}$ is initially filled
with a 0.1~mol/L solution of a solute which cannot permeate the
membrane, then transferred into a 0.105~mol/L solution of a solute
which can permeate the membrane ($P_s = 2\cdot
10^{-8}\;\textrm{m/s}$).  Other parameters used were $\bar{V}_w =
1.806\cdot 10^{-2}\;\textrm{L/mol}$, $P_f = 2.23\cdot
10^{-4}\;\textrm{m/s}$. Upon transfer into a hypertonic solution, two
processes are competing: the efflux of water, driven by the osmotic
pressure mismatch, is the quicker of the two, while the influx of the
solute, accompanied by the concomitant influx of water, is the slower
one.  Therefore, we first have to solve the system defined by
\refeqn{eq:dV-2p-flaccid} and \refeqn{eq:dNs-2p-flaccid}, with $V(0) =
V_0$ and $N_s(0) = 0$ as initial conditions.  A deflated vesicle
deforms into an approximate oblate spheroid; since in the experimental
setup, the optical axis is usually aligned with the symmetry axis of
the spheroid, the cross-section radius increases with the decreasing
volume.  Introducing $v = V/V_0$ and $x = R_1/R_0$, $R_1$ being the
spheroid visible semiaxis, one obtains the following relationship
between $v$ and $x$:
\begin{equation}
2 x^2 + \frac{v^2}{x^4 \sqrt{1-v^2/x^6}} \ln\left( 
  \frac{1 + \sqrt{1-v^2/x^6}}{1 - \sqrt{1-v^2/x^6}} \right) = 4 \; .
\label{eq:oblate-spheroid}
\end{equation}
After a certain time, the vesicle becomes spherical again. From this
point onwards, it starts to grow while maintaining a spherical shape,
and its behaviour is governed by \refeqn{eq:dV-2p-sph} and
\refeqn{eq:dNs-2p-sph}, with initial conditions $R(0) = R_0$ and
$c_s^i(0)$ equal to the concentration of the permeable solute, which
diffused into the vesicle during the previous step.

\subsection{One-parameter model}

Vesicle swelling is a two step process: in the first step, permeating
solute diffuses into vesicle interior and thus causes osmotic
non-equilibrium, and in the second step (which constitutes the bulk of
swelling), it is followed by the water influx, which balances the
osmotic non-equilibrium.  \refeqn{eq:dV-2p-flaccid} and
\refeqn{eq:dNs-2p-flaccid} can be regarded as a kinetic system.  In
the systems treated here, however, the influx of water is a much
faster process than the solute diffusion, and thus
\refeqn{eq:dNs-2p-flaccid} represents the rate-limiting step.  In the
case where solute diffusion is a much slower process than the influx
of water, a simpler description can be obtained.

In this simpler description, \refeqn{eq:dNs-2p-flaccid} is retained,
while \refeqn{eq:dV-2p-flaccid} is replaced by an instantaneous
osmotic equilibrium:
\begin{equation}
  \label{eq:osm-equilib}
  c_s^e = \frac{N_s + N_n}{V} \; .
\end{equation}
With $c_s^e$ and $N_n$ being constant, this leads to
$\mathrm{d}N_s/\mathrm{d}t = c_s^e\mathrm{d}V/\mathrm{d}t$.  Taking
this into account, and inserting \refeqn{eq:osm-equilib} into
\refeqn{eq:dNs-2p-flaccid}, one obtains
\begin{equation}
  \label{eq:dV-1p}
  \frac{\mathrm{d}V}{\mathrm{d}t} = \frac{P_s A}{c_s^e} \frac{N_n}{V} \; .
\end{equation}
\refeqn{eq:dV-1p} is general in the case that both $A$ and $V$ may
vary with time.  Again, two special cases can be considered: flaccid
vesicles, where $A = \textrm{const.}$, and spherical vesicles, where
both $A$ and $V$ can be expressed in terms of the vesicle radius $R$.
In the latter case, a further simplification is possible by assuming
that initially, the vesicle was spherical and in osmotic equilibrium,
\emph{i.e.}, $c_s^e = N_n/(4\pi R_0^3/3)$.  In this case,
\refeqn{eq:dV-1p} can be rewritten into
\begin{equation}
  \label{eq:dr-1p}
  \frac{\mathrm{d}R}{\mathrm{d}t} = P_s \left(\frac{R_0}{R}\right)^3 \; ,
\end{equation}
which can be readily integrated:
\begin{equation}
  \label{eq:r-1p}
  R(t) = R_0 \left(1 + \frac{4 P_s t}{R_0}\right)^{1/4} \; .
\end{equation}
As long as deformations are small, $R$ exhibits linear growth, $R(t)
\approx R_0 + P_s t$.  In this model, which assumes that the membrane
permeability for water exceeds the permeability for solute by such a
large margin that water transport can be considered instantaneous, the
only parameter governing the vesicle volume change is $P_s$.  Despite
its simplifications, the one-parameter model has proved to be useful
in certain situations \cite{Mazur:1974,Mazur:1976}.

\subsection{The influence of membrane elasticity}

The treatment of osmotic swelling of spherical vesicles presented so
far assumes that the membrane is infinitely ``soft'' and does not
oppose its stretching as the vesicle swells.  In reality, the
stretching modulus of a phospholipid membrane is finite.  This means
that the apparent value of membrane permeability derived from the
experiments with osmotic swelling of spherical vesicles is slightly
underestimated: because the vesicle membrane opposes its stretching,
the radius increases slightly less in a given interval of time than it
would if the limiting factor was the membrane permeability alone.

It is common to assume that the membrane area has a certain relaxed
area $A_0$, and expand the free energy of membrane stretching around
this value:
\begin{equation}
  \label{eq:stretching-energy}
  W = \frac{K}{2 A_0} (A - A_0)^2 \; .
\end{equation}
Here, $K$ is the membrane stretching modulus, and $A_0 = 4\pi R_0^2$.
We consider now the work $p\,\mathrm{d}V$, needed to increase the
radius of a spherical vesicle by a small amount $\mathrm{d}R$.  The
corresponding change in the free energy is
\[ \mathrm{d}W = K \frac{A - A_0}{A_0} \mathrm{d}A \; .
\]
Substituting $\mathrm{d}V = 4\pi R^2 \mathrm{d}R$, $\mathrm{d}A = 8\pi
R\, \mathrm{d}R$, and equating $p\,\mathrm{d}V$ with the change in the
free energy, one obtains:
\begin{equation}
  \label{eq:pressure}
  p = \frac{2 K}{R} \frac{R^2 - R_0^2}{R_0^2} \; .
\end{equation}
In the case of osmotic equilibrium, this pressure is balancing the
difference in the osmotic pressure: $p = \Delta\Pi$.

When the pressure exerted by the membrane is taken into account,
\refeqn{eq:dV-2p-flaccid} transforms into
\begin{equation}
  \label{eq:dV-2p-pressure}
  \frac{\mathrm{d}V}{\mathrm{d}t} = - L_p A (\Delta\Pi - \Delta p) \; .
\end{equation}
Here, $\Delta\Pi = \alpha(c^e - c^i)$ and $\Delta p = p^e - p^i = 0 - p$,
where $c^e = c_s^e$, $c^i = c_n^i + c_s^i$, and $p$ is defined by
\refeqn{eq:pressure}.  In the system defined by \refeqn{eq:dV-2p-sph}
and \refeqn{eq:dNs-2p-sph}, the vesicle membrane would in theory grow
indefinitely because the osmotic pressure is never entirely balanced.
A membrane which opposes vesicle swelling limits the extent of
swelling by the condition $\Delta\Pi = \Delta p$.

\begin{figure}[t]
  \centering\includegraphics{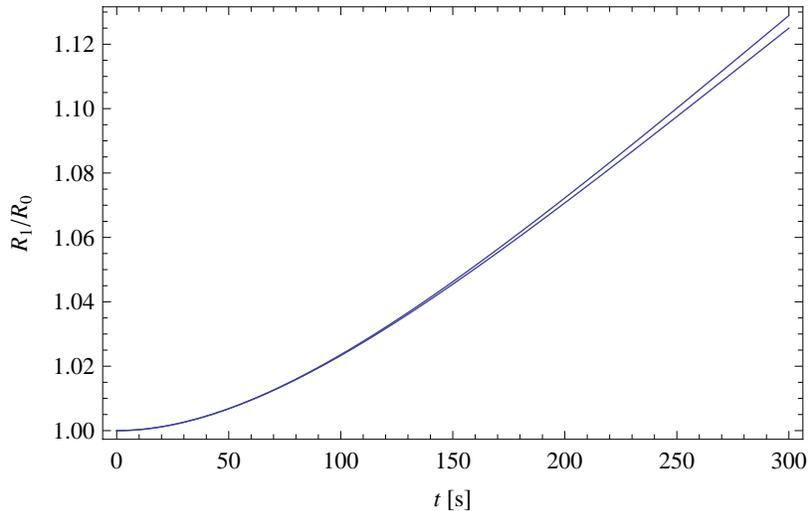}
  \caption{The radius of a vesicle cross-section upon transfer into an
    isotonic solution of a solute which can permeate vesicle membrane,
    as calculated by the two-parameter model.  The lower curve shows
    the rate of vesicle growth when the finite membrane stretching
    modulus $K$ is taken into account, while the upper curve shows the
    dependency for $K=0$. Calculation parameters are given in the
    text.}
  \label{fig:membrane-influence-2p}
\end{figure}

Rewriting \refeqn{eq:dV-2p-pressure} for a spherical vesicle, one
obtains an equation analogous to \refeqn{eq:dNs-2p-sph}:
\begin{equation}
  \label{eq:dr-2p-sph-membrane}
  \frac{\mathrm{d}R}{\mathrm{d}t} = -L_p \left[ \alpha
    \left( c_s^e - c_s^i - c_{n0}^i \left(\frac{R_0}{R}\right)^3 \right)
    + \frac{2K}{r} \left(\left(\frac{R}{R_0}\right)^2 - 1\right) \right] \; .
\end{equation}
For realistic parameter values, membrane stretching only adds a minor
correction to the vesicle swelling rate.
Figure~\ref{fig:membrane-influence-2p} shows the swelling of a
spherical vesicle upon transfer into an isotonic medium of permeating
solute ($P_s = 2\cdot 10^{-8}\;\textrm{m/s}$), obtained by the
numerical solution of the two-parameter system defined by
\refeqn{eq:dr-2p-sph-membrane} and \refeqn{eq:dNs-2p-sph}, with $R(0)
= R_0$ and $c_s^i(0) = 0$ as initial conditions.  The upper curve
disregards the membrane stretching energy ($K = 0$), while the lower
uses the value $K = 0.23\;\textrm{N}\,\textrm{m}^{-1}$. Other
parameters used in calculation are $c_s^e = c_{n0}^i =
0.1\;\textrm{mol}\,\textrm{L}^{-1}$, $R_0 = 20\;\upmu\textrm{m}$, $L_p
= 1.645\cdot 10^{-13}\;\textrm{m}\,\textrm{s}^{-1}\,
\textrm{Pa}^{-1}$, and $\alpha = RT = 2477\;
\textrm{J}\,\textrm{mol}^{-1}$ at $T = 298\;\textrm{K}$.

\subsection{Swelling-burst cycle}

In reality, vesicles approximately follow the swelling course shown in
Figure~\ref{fig:membrane-influence-2p} only until the critical strain
of the membrane is reached.  Once the critical strain is reached, the
membrane ruptures, and the vesicle bursts. Subsequently, the membrane
is resealed and another cycle of swelling commences.  Experimentally,
long trains of $\sim 50$ vesicle bursts have been observed.

The phenomenon has been predicted and thoroughly worked out from the
theoretical point of view by Kozlov and Markin \cite{Koslov:1984};
apparently unaware of their work, other authors attempted the same
decades later \cite{Popescu:2008}.  Experimentally, the effect has
been observed on erythrocytes when their suspending medium has been
exchanged with a hypotonic one \cite{ZadeOppen:1998}, on giant
adhering vesicles \cite{Sandre:1999}, on giant vesicles with
equinatoxin-II-induced pores \cite{Mally:2002}, on giant vesicles with
photoinduced pores \cite{Karatekin:2003a}, on giant vesicles with
mellitin-induced pores \cite{Mally:2007}, and on giant vesicles upon
transfer from a solution of a non-permeating solute to a solution of a
permeating solute \cite{Peterlin:2008a}.

Kozlov and Markin \cite{Koslov:1984} consider a vesicle filled with an
osmotically active solute (\emph{i.e.}, non-permeating), and write an
expression for the osmotic influx $J_\mathrm{inf}$ of water:
\begin{equation}
  \label{eq:Jinf}
  J_\mathrm{inf} = - L_p A (\Delta\Pi - \Delta p) \; ,
\end{equation}
where $L_p$ is the permeability of membrane for water (hydraulic
conductivity), $A$ is the membrane area, $\Delta\Pi = \Pi^e - \Pi^i$
is the difference in osmotic pressure, and $\Delta p = p^e - p^i$ is
the difference in the hydraulic pressure.  \refeqn{eq:Jinf} is
identical to \refeqn{eq:dV-2p-pressure} above.

The vesicle is treated as spherical, its radius denoted by $R$.  Due to
osmotic swelling, $R > R_0$, where $R_0$ denotes the value of a
non-expanded vesicle.  Denoting $R = R_0 + \Delta R$, the
corresponding changes in the vesicle volume and membrane area are
$\Delta A/A_0 \approx 2 \Delta R/R_0$ and $\Delta V/V_0 \approx 3
\Delta R/R_0$, respectively. A stretched membrane creates a pressure
inside the vesicle -- cf. \refeqn{eq:pressure}, which opposes the swelling,
\[ \Delta p = \frac{2 K}{R} \frac{\Delta A}{A_0} \; ,
\]
with $K$ being the membrane stretching modulus.  The elastic free
energy associated with membrane stretching is given according to
\refeqn{eq:stretching-energy}:
\[ W_a = \frac{K}{2} \frac{(\Delta A)^2}{A_0} \; .
\]
A pore present in the membrane lowers $\Delta A$ and thus reduces this
energy term.  However, on the other hand, creating a tension pore
increases the total energy by the edge energy, equal to the product of
the lenght of the pore edge, $2 \pi r$, and the linear tension
$\gamma$:
\begin{equation}
  \label{eq:edge-energy}
  W_p = 2\pi \gamma r
\end{equation}
Here, $r$ refers to the radius of a circular tension pore.
\refeqn{eq:edge-energy} is an approximate expression, valid for large
pores.  A consequence of a pore having a non-zero energy is that the
intravesicular pressure does not drop to zero after the membrane
ruptures.  Instead, a residual intravesicular pressure remains:
\begin{equation}
  \label{eq:residual-pressure}
  \Delta p_r = \frac{2\gamma}{R_0 r} \; .
\end{equation}
This residual intravesiculare pressure causes the efflux of the
intravesicular solution once the vesicle ruptures.  The efflux
$J_\mathrm{eff}$ is approximately equal to
\begin{equation}
  \label{eq:Jeff}
  J_\mathrm{eff} = \frac{r^3 \Delta p_r}{\eta} \; .
\end{equation}
Here $\eta$ denotes the viscosity.

\begin{figure}[t]
  \centering\includegraphics{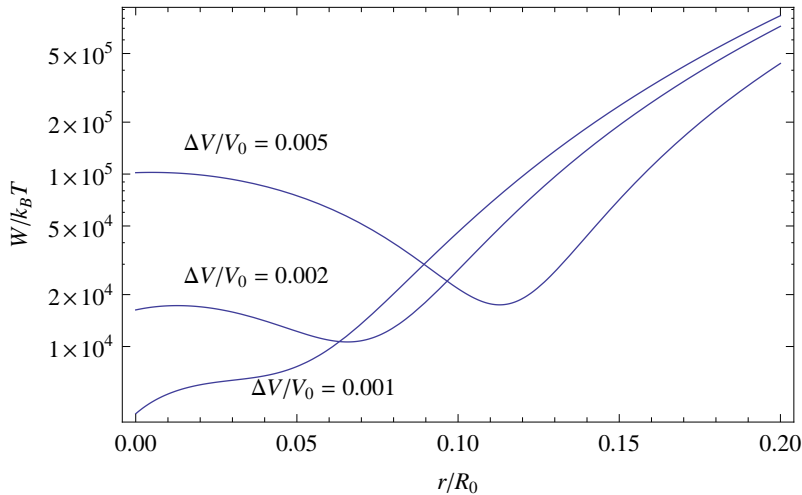}
  \caption{The elastic energy of a vesicle with a pore plotted against
    reduced pore radius $r/R_0$ for several different values of $\Delta
    V/V_0$.  Vesicle elastic energy is given by
    \protect\refeqn{eq:Wsum}, with $R_0 = 5\;\upmu\textrm{m}$, $K =
    240\;\textrm{mN/m}$, $\gamma = 20\;\textrm{pN}$.}
  \label{fig:pore-energy}
\end{figure}

Substituting \refeqn{eq:residual-pressure} into
\refeqn{eq:dV-2p-pressure} and integrating it, one obtains the time
course of vesicle volume upon transfer into a hypotonic medium:
\begin{equation}
  \label{eq:volume}
  \Delta V(t) = \frac{\pi R_0^4 \alpha\,\Delta c_n}{K}
  \left[ 1 - \exp\left(-\frac{4 K L_p}{R_0^2}\:t\right) \right] \; .
\end{equation}
Here, $\alpha = RT$. The elastic energy of a vesicle with a pore is
the sum of the membrane stretching energy and the linear energy of the
pore:
\begin{equation}
  \label{eq:Wsum}
  W = \frac{K}{2} \frac{(\Delta A - A_p)^2}{A_0} + 
  2 \sqrt{\pi} \gamma \sqrt{A} \; .
\end{equation}
At a given $\Delta A$, the area of the tension pore $A_p$ adapts in
such a way as to minimize the total energy given by \refeqn{eq:Wsum}.
If $W$ is plotted against pore radius $r$ for realistic values of $K$
and $\gamma$ and for different values of $\Delta V/V_0$, one can prove
that for low values of $\Delta V/V_0$, $W$ is a monotonously
increasing function of pore radius $r$ (figure \ref{fig:pore-energy}).
If $\Delta V/V_0$ is increased, we reach a critical value
$\Delta\tilde{V}/V_0$ at which $W(r)$ has an inflection point. By
increasing $\Delta V/V_0$ even further, a local minimum becomes a
global one, meaning that at some non-zero pore radius, the energy of
the vesicle is lower than at $r = 0$, and by overcoming the energy
barrier, the vesicle can jump from a poreless state into a state with
a pore.

Kozlov and Markin show that based on the difference between the
influx and the efflux, given by \refeqn{eq:Jinf} and \refeqn{eq:Jeff},
three different scenarios are possible:
\begin{quote}
  \begin{tabular}{rl}
    $J_\mathrm{inf} > J_\mathrm{eff}$ & instant efflux of the whole 
    vesicle interior \\
    $J_\mathrm{inf} = J_\mathrm{eff}$ & steadily open pore \\
    $J_\mathrm{inf} < J_\mathrm{eff}$ & pulse-wise regime
  \end{tabular}
\end{quote}
Kozlov and Markin conclude that for realistic parameter values, the
pulse-wise regime is the most probable one, which is consistent with
our own observations.

In the pulse-wise regime, it is possible to infer some relationships
between relevant quantities during successive bursts by employing two
assumptions:
\begin{enumerate}
\item During the swelling phase, the amount of the non-permeating
  solute inside the vesicle remains constant.
\item At burst, partial concentrations of both the permeating and the
  non-permeating solute remain constant.
\end{enumerate}
The first assumption is valid if the characteristic time for solute
exchange is long compared with the time course of the experiment; the
second assumption assumes that the interior of the vesicle is well
mixed.

\begin{figure}[t]
  \centering\includegraphics[width=0.9\linewidth]{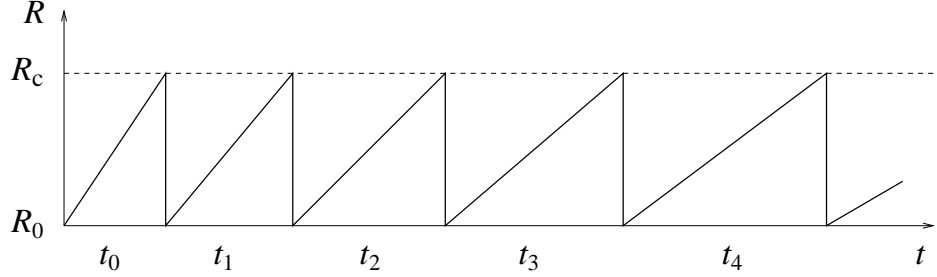}
  \caption{In the pulse-wise regime, vesicle radius $R$ increases due
    to osmotic swelling from its relaxed value $R_0$ until it reaches
    its critical value $R_c$, at which the critical strain for the
    membrane is reached. At that point, membrane ruptures, vesicle
    ejects part of its interior, upon which the membrane reseals and
    the radius returns to its relaxed value.}
  \label{fig:burst-train}
\end{figure}

While the validity of the above assumptions is more general, some
simple formulas can be obtained for the linearized one-parameter case.
Let us consider a vesicle in the pulse-wise regime (figure
\ref{fig:burst-train}).  Due to osmotic swelling, the vesicle radius
$R$ increases from its relaxed value $R_0$ until a critical value
$R_c$ is reached, at which the critical strain for the membrane is
reached, upon which membrane ruptures, ejects a part of the internal
volume, and reseals in a relaxed state.  As the $(A_c - A_0)/A_0$
amounts to a few percent, \refeqn{eq:r-1p} can be linearized.

\refeqn{eq:r-1p} is valid only up to the first burst.  Deriving it, we
assumed $c_s^e = N_n/V_0$.  We can derive the concentration of
non-permeating solute in subsequent swelling cycles using the
assumptions above. Denoting the amount of non-permeating solute inside
the vesicle after $n$ bursts by $N_n^{(n)}$, the concentration of
non-permeating solute inside the vesicle after $n$ bursts by
$c_n^{(n)}$, $V_0 = 4\pi R_0^3/3$, and $V_c = 4\pi R_c^3/3$, one can write:
\begin{alignat*}{3}
  N_n^{(0)} &= c_n^{(0)} V_0 \;&= c_n^{(1)} V_c \\
  N_n^{(1)} &= c_n^{(1)} V_0 \;&= c_n^{(2)} V_c \\
   & \qquad \ldots &
\end{alignat*}
A general expression for $c_n^{(n)}$ is then
\begin{equation}
  \label{eq:cn(n)}
  c_n^{(n)} = c_n^{(0)} \left(\frac{V_0}{V_c}\right)^n \; .
\end{equation}
In our experimental setup, $c_n^{(0)} = c_s^e$ holds.  Substituting
$c_n^i$ with $c_n^{(n)}$ in the derivation of \refeqn{eq:r-1p}, one
obtains the temporal dependence of vesicle radius after $n$ bursts:
\begin{equation}
  \label{eq:rn-1p}
  R^{(n)}(t) = R_0\left[ 1 + \left(\frac{V_0}{V_c}\right)^n 
    \frac{4 P_s t}{R_0}\right]^{1/4} \; .
\end{equation}
Or, linearized, $R^{(n)}(t) = R_0 + (V_0/V_c)^n P_s t$. The time
between two successive bursts is then
\[ t_n = \left(\frac{V_0}{V_c}\right)^n \frac{R_c - R_0}{P_s} \; .
\]
Denoting $\alpha_c = (A_c - A_0)/A_0$ and expressing $V_c/V_0 \approx
1 + (3/2)\alpha_c$, $R_c/R_0 \approx 1 + (1/2)\alpha_c$, one obtains
after linearization for the time elapsed between the $n$-th and the
$n+1$-th burst:
\begin{equation}
  \label{eq:tn-1p}
  t_n \approx \frac{R_0}{P_s} \left(1 + \frac{\alpha_c}{2} + 
    \frac{3\alpha_c}{2} n \right) \;.
\end{equation}
In this linear model, the time elapsed between subsequent bursts
increases linearly with $n$.

As the formation of a tension pore is a stochastic event, $R_c$ is not
a constant, but can vary from one burst to another.  To account for
this, \refeqn{eq:cn(n)} can be written as
\begin{equation}
  \label{eq:cn-var}
  c_n^{(n)} = c_n^{(0)} \prod_{j=0}^{n-1} \frac{V_0}{V_c^{(j)}} \; ,
\end{equation}
where the index $j$ denotes each of the $n$ bursts.  We have
introduced $V_c^{(j)} = 4\pi R_c^{(j)}/3$, the volume of vesicle at
which the critical strain for the membrane is reached in the $j$-th
burst.

\subsection{Critical phenomena in osmotic swelling}

Upon transfer of vesicles from a solution of a non-permeating solute
into a solution of a permeating solute, vesicles often exhibit a
transient increase of cross-section radius (figure
\ref{fig:burst-seq}).  We attribute this behaviour to a slight
hypertonicity of the target solution, which causes the vesicles to
slightly deflate and change their shapes from an initially spherical
shape into an oblate spheroid.  The concentration gradient, however,
drives the permeating solute into the vesicle interior, and the
ensuing osmotic pressure difference serves as a driving force for
water, which causes the vesicle to re-inflate.  The inflation process
has two stages: initially, the vesicle is flaccid and approaching the
spherical shape, while in the second stage, the vesicle is spherical,
its membrane being increasingly stretched.

It has been shown that an inflating vesicle reaches the end of the
first stage critically, through a continous transition.  This was
initially demonstrated for unstretchable membrane \cite{Haleva:2008a},
and later the theory has been extended in order to account for
membrane stretching \cite{Peterlin:2012}.  We shall provide a brief
outline of the underlying theory, which employs a thermodynamic
framework \cite{Diamant:2011}.

Let us consider a vesicle with $Q$ entrapped molecules of a
non-permeating solute.  We assume that the vesicle is brought into a
thermal equilibrium with the surroundings, characterized by
temperature $T$ and pressure $p_0$. Both the vesicle volume $V$ and
the surface area $A$ are treated as free, independent thermodynamic
variables.  The Gibbs free energy of the system contains volume and
surface contributions:
\begin{equation}
  \label{eq:Gibbs}
  G = G_\mathrm{3D} + F_\mathrm{2D} \; .
\end{equation}
Considering the enclosed solution as ideal and dilute, one can write
\begin{equation}
  \label{eq:G_3D}
  G_\mathrm{3D} = k_B T Q \left(\ln\frac{Q\bar{V}_w}{V} - 1\right) + p_0 V \; .
\end{equation}
The surface terms comprise the stretching and the bending parts,
\begin{equation}
  \label{eq:F_2D}
  F_\mathrm{2D} = F_s + F_b \; .
\end{equation}
The stretching part is assumed to have a minimum at some value $A_0$;
a quadratic  expansion is used:
\[ F_s = \frac{K}{2} \frac{(A - A_0)^2}{A_0} \; .
\]
In regarding the bending free energy, only the contribution from
undulation entropy is included, which describes the suppression of
bending fluctuations as the vesicle shape approaches a sphere.
Assuming small fluctuations around a spherical shape, it is given by
\cite{Haleva:2008a}
\begin{equation}
  \label{eq:F_b}
  F_b = - \frac{N}{2} k_B T \,\ln(1 - v) \; .
\end{equation}
Here, $N$ is the total number of independent bending modes
contributing to the membrane thermodynamics, and $v$ is the vesicle
reduced volume.  It is important to note that through the reduced
volume (which depends on both volume and surface area), $F_b$ provides
the coupling between the volume part $G_\mathrm{3D}$ and the surface
part $F_\mathrm{2D}$.  The number of modes $N$ is not known in advance
and is a parameter extracted from the experiment.

Using Eqs.~\ref{eq:Gibbs}--\ref{eq:F_b}, the Gibbs free energy of a
vesicle is defined via $(T, p_0, Q, N)$, as well as $A$ and $V$.  By
minimizing $G$ with respect to $A$ and $V$ the equilibrium free energy
can be obtained.  Introducing an intensive area $a = A/N$ and its
relaxed value $a_0 = A_0/N$ along the way, one obtains a set of
equations from which $a$ and $v$ can be calculated:
\begin{align}
  \frac{1}{2} \delta_N \frac{v}{1-v} + \left(\frac{a}{a_0}\right)^{3/2} v &
  = \frac{Q}{Q_c} \; , \label{eq:Haim8} \\
  \frac{2}{3} \frac{\delta_N}{\delta_K} \frac{a(a-a_0)}{a_0^2} + 
  \left(\frac{a}{a_0}\right)^{3/2} v &= \frac{Q}{Q_c} \; . \label{eq:Haim9}
\end{align}
In \refeqn{eq:Haim8} and \refeqn{eq:Haim9}, we have defined
\begin{equation}
  \label{eq:Qc}
  Q_c = \frac{p_0 V_0}{k_B T} \; , \quad \text{where} \quad 
  V_0 = \frac{A_0^{3/2}}{6\sqrt{\pi}} \; ,
\end{equation}
and parameters $\delta_N$ and $\delta_K$, the first one related to the
finite vesicle size, and the second one to the finite membrane
stretchability:
\begin{alignat}{4}
  \delta_N &=&\; \frac{N}{Q_c} &\propto N^{-1/2} \; , \\
  \delta_K &=&\; \frac{k_B T}{K a_0} &\propto K^{-1} \; .
\end{alignat}

Two limiting cases of the described systen can be considered.  Setting
$N \rightarrow \infty$, $\delta_N \rightarrow 0$ while keeping the
stretching modulus finite brings us to the usual thermodynamic limit
of an infinite system. In this limit, \refeqn{eq:Haim8} and
\refeqn{eq:Haim9} are degenerate and yield the expected equilibration
of the internal and the external pressure: $Q k_B T / V = p_0$.  The
second limiting case corresponds to an unstretchable membrane: $K
\rightarrow \infty$, $\delta_K \rightarrow 0$.  In this case,
\refeqn{eq:Haim9} requires $a = a_0$, while \refeqn{eq:Haim8} yields
the behaviour studied in Ref.~\cite{Haleva:2008a}, with $v$ exhibiting a
critical behaviour as $N \rightarrow \infty$.

Away from either limit, \refeqn{eq:Haim8} and \refeqn{eq:Haim9} yield
\begin{equation}
  \label{eq:Haim12}
  \delta_K \frac{v}{1 - v} = \frac{4}{3} \frac{a(a-a_0)}{a_0^2} \; .
\end{equation}
We can see from \refeqn{eq:Haim12} that attaining a perfectly
spherical shape ($v = 1$) would reqire infinite strain $a \rightarrow
\infty$, which is not realistic.  The deviation of the area from its
relaxed state, $a - a_0$, is inversely proportional to the deviation
of the vesicle shape from a sphere, $1 - v$.  As $\delta_K \ll 1$, it
takes a highly swollen vesicle, $1 - v \sim \delta_K$, to produce a
significant membrane strain $(a - a_0)/a_0$.

In order to provide an adequate description of the transition, we
define the control parameter $q = Q/Q_c-1$, proportional to the number
of enclosed molecules, and order parameter $M = 1 - v$, which serves
as a measure of the deviation of vesicle shape from a sphere. Solving
\refeqn{eq:Haim8} and \refeqn{eq:Haim9} for $M$ in the vicinity of the
transition, one obtains
\begin{equation}
  \label{eq:Haim14}
  M(q) = \frac{\Delta}{2}\left( \sqrt{1 + \left(\frac{q}{\Delta}\right)^2}
    - \frac{q}{\Delta}\right) ; \qquad 
  \Delta = \sqrt{2 \delta_N + \frac{9}{2} \delta_K} \; .
\end{equation}
For $q < 0$, $M(q) \simeq |q|$, which is appreciable, while for $q >
0$, $M(q) \sim \Delta/(4 q)$, which is very small.  The transition
occurs over a region, determined by $\Delta$.  Thus, both the finite
vesicle size and the membrane stretchability contribute to widening of
the transition. In the limit $\Delta \rightarrow 0$, \emph{i.e.}, for
an infinitely large vesicle enclosed by a non-stretchable membrane,
$\mathrm{d}M/\mathrm{d}q$ has a discontinuity at $(q = 0, M = 0)$.
Setting $\delta_K = 0$ while keeping $\delta_N$ finite yields back the
results obtained in \cite{Haleva:2008a}.  Which of the two factors
entering $\Delta$ is dominant, $\delta_N$ or $\delta_K$, depends on
their ratio
\[ \frac{\delta_N}{\delta_K} \sim \frac{K/R_0}{p_0} \; .
\]
Using the values from our experiments -- $R_0 \sim
20$--$50\;\upmu\textrm{m}$, $c_0 \sim 0.1$--$0.2\;\textrm{mol/L}$ --
as well as $K = 240\;\textrm{mN/m}$ \cite{Rawicz:2000}, we obtain
$\delta_N/\delta_K \sim 0.01$--0.1. Therefore, in our conditions, the
transition width is governed by the finite stretching modulus, and can
be approximated by
\begin{equation}
  \label{eq:Haim18}
  \Delta \approx \left(\frac{9}{2} \delta_K\right)^{1/2} =
  \left(\frac{9 k_B T}{2 K a_0} \right)^{1/2} \; .
\end{equation}

While \refeqn{eq:Haim14} fully describes the law of corresponding
states for osmotic swelling of nearly spherical vesicles, we would
like to transform it into a form which would allow us to compare it
with the experiment, \emph{i.e.,} with the function $R(t)$.

The simpler part is the transformation of the time axis.  As the
concentration of the permeating solute inside the vesicle is
approximately two orders of magnitude lower than its concentration in
the surrounding medium, the former can be neglected, yielding
$\mathrm{d}Q/\mathrm{d}t \approx P A_0 c_0$, which further leads to
\begin{equation}
  \label{eq:Haim19}
  q = (3 P/R_0) t + \text{const.}
\end{equation}
\refeqn{eq:Haim19} gives a simple linear relationship between the
time $t$ and the control parameter $q$.

The transformation for the radius is less straightforward.  Here we
only reproduce the result, which is worked out in detail in
\cite{Peterlin:2012}.  Denoting $g_\mathrm{exp} = (R_1(t)/R_0)^3 - 1$,
where $R_1(t)$ is the radius of vesicle projected cross-section, one
arrives at the following scale and shift transformation,
\begin{equation}
  \label{eq:Haim20}
  f(\xi) = \sqrt{\frac{8}{15 \Delta}} \left[ g_\mathrm{exp} \left(
      \xi - \frac{1}{4} \left(\frac{15}{\Delta}\right)^{1/3}\right) - \xi\Delta 
      + \frac{3}{4} \left( 15 \Delta^2 \right)^{1/3} \right] \; ,
\end{equation}
where $\xi = q/\Delta$ and $q = (3 P/R_0) t$.  Using the permeability
$P$ and the transition width $\Delta$ as two fitting parameters, one
can verify that the experimental data for all recorded vesicle
transfers collapse onto a single universal function,
\begin{equation}
  \label{eq:Haim21}
  f(\xi) = \left( \sqrt{1 + \xi^2} - \xi\right)^{1/2} \; .
\end{equation}
Fitting \refeqn{eq:Haim21} to the recorded data $R_1(t)$ of vesicle
transfer into an isotonic solution of urea yields the permeability of
membrane for urea, $P_s = 0.013 \pm 0.001\;\upmu\textrm{m/s}$.  The
obtained permeabilities for glycerol and ethylene glycol, on the other
hand, exhibit concentration dependence.  The values obtained for
glycerol range from 0.0053 at $c_0 = 0.1$~M to $0.019 \pm
0.006\;\upmu\textrm{m/s}$ at 0.2~M, and those for ethylene glycol
range from $0.046 \pm 0.006\;\upmu\textrm{m/s}$ at 0.1~M to 
$0.085 \pm 0.01\;\upmu\textrm{m/s}$ at 0.2~M.

\section{Experimental analysis}

In this section, we will present a method for determining membrane
permeability for a given solute from the analysis of swelling-burst
cycles \cite{Peterlin:2009a}, and demonstrate that its results are
consistent with the analysis of the osmotic swelling of flaccid
vesicles \cite{Peterlin:2012}.

In section \ref{sec:theory}, \refeqn{eq:rn-1p} has been obtained,
which, when linearized, shows that before the first burst, the slope
of $R(t)$ equals to the permeability $P_s$, and after $n$ bursts, the
slope becomes $(V_0/V_c)^n P_s$ if all bursts are equal, and $P_s
\prod_{j=0}^{n-1} (V_0/V_c^{(j)})$ if they are not.  The underlying
assumptions are that the vesicle is spherical at the onset of the
first swelling cycle, and the concentration of the non-permeating
solute in its interior is equal to the concentration of the permeating
solute outside.

These assumptions are not always met.  In 37 out of 47 recorded
vesicle transfers in our experiments, the dependence of the vesicle
cross-section radius as a function of time exhibits an initial
transient maximum or ``bulge'' like the one seen in
Figure~\ref{fig:burst-seq}.  In those cases, the vesicle is filled
with a mixture of a non-permeating and permeating solute already at
the onset of the first swelling cycle, and \refeqn{eq:rn-1p} needs a
small correction to take that into account.

An estimate of the mole fraction of the non-permeating and the
permeating solute at the onset of the first swelling cycle can be
obtained from the initial transient maximum.  If we approximate the
shape of a partially deflated vesicle with an oblate spheroid and
denote $x = R_1/R_0$, where $R_1$ is the radius of the vesicle cross
section at the crest of the ``bulge'' and $R_0$ is its value in the
``valley'', \emph{i.e.}, at the onset of the first swelling cycle, we
can compute the reduced volume of the vesicle ($v$) using
\refeqn{eq:oblate-spheroid}.

We can show that the reduced volume of the vesicle is equal to the
mole fraction of the osmotically active solute at the onset of the
first swelling cycle.  At the crest of the bulge, the vesicle is
flaccid with a volume $V = v V_0 < V_0$, and containing $N_\mathrm{n}$
moles of non-permeating solute. At the onset of the first swelling
cycle, the vesicle is spherical with a radius $R_0$, and containing
$N_\mathrm{n}$ moles of non-permeating solute and $N_\mathrm{s}$ moles
of permeating solute.  As the rapid osmotic exchange of water ensures
that the total concentration is equal in both cases, $N_\mathrm{n}/V =
(N_\mathrm{n} + N_\mathrm{s})/V_0$, it follows
\[ v = \frac{N_\mathrm{n}}{N_\mathrm{n} + N_\mathrm{s}} = x_\mathrm{n} \; .
\]
We can amend \refeqn{eq:rn-1p} with this correction, yielding
\begin{equation}
  \label{eq:rnx-1p}
  R^{(n)}(t) = R_0\left[ 1 + \left(\frac{V_0}{V_c}\right)^n 
    \frac{4 x_n P_s t}{R_0}\right]^{1/4} \; .
\end{equation}
We need to note that taking the value of $x_n$ at the onset of a
swelling cycle is an approximation, as $x_n$ actually changes during
the course of a swelling cycle.

Treating the vesicle swelling-burst cycles as piecewise linear, the
slope of the $k$-th cycle can be expressed as
\begin{equation}
  \label{eq:slope-exp}
  \frac{\Delta R^{(k)}}{\Delta t^{(k)}} = x_\mathrm{n} P_\mathrm{s}^{(k)} 
  \prod_{j = 0}^{k - 1} \frac{V_0}{V_\mathrm{c}^{(j)}} \; .
\end{equation}
While membrane permeability for the permeating solute can be in
principle computed from an individual vesicle-swelling phase between
two bursts, a more reliable estimate is obtained by averaging it over
all $n$ bursts in the cycle:
\begin{equation}
  \label{eq:permeability-cycle}
  \bar{P}_\mathrm{s} = \frac{1}{n} \sum_{k = 1}^n \frac{1}{x_\mathrm{n}}
  \frac{\Delta R^{(k)}}{\Delta t^{(k)}} 
  \prod_{j = 0}^{k - 1} \frac{V_0}{V_\mathrm{c}^{(j)}} \; .
\end{equation}
\refeqn{eq:permeability-cycle} gives the estimate for the membrane
permeability based on a single recording of a vesicle transfer.
Averaging over several recordings of vesicle tranfer are needed to
obtain a more reliable estimate of membrane permeability.

Table~\ref{tab:permeability} summarizes the permeability data for
glycerol, urea, and ethylene glycol, obtained by using two different
analyses.  The term ``burst train'' refers to the analysis presented
here, while the term ``critical swelling'' refers to the alternative
analysis presented in the preceding section \cite{Peterlin:2012},
where vesicle transition from a flaccid to a spherical state before
the first burst is observed.

\renewcommand{\arraystretch}{1.2}
\begin{table}
  \caption{Permeability estimates for glycerol, urea, and ethylene glycol, 
    determined from series of micrographs of giant unilamellar 
    vesicles upon transfer from a sucrose/glucose solution into an 
    isomolar solution of a given solute. ``Burst train'' refers to the
    analysis presented here, and ``critical swelling'' to an alternative
    analysis \cite{Peterlin:2012} on the same data set.}
  \begin{center}
    \begin{tabular}{ll@{$\pm$}ll@{$\pm$}ll@{$\pm$}ll@{$\pm$}l}
      \hline
      & \multicolumn{8}{c}{Permeability $P_\mathrm{s}$ [$\upmu$m/s]} \\
      solute & \multicolumn{4}{c}{``burst train''}
      & \multicolumn{4}{c}{``critical swelling''} \\
      & \multicolumn{2}{c}{0.1~M} & \multicolumn{2}{c}{0.2~M} 
      & \multicolumn{2}{c}{0.1~M} & \multicolumn{2}{c}{0.2~M} \\
      \hline
      glycerol & 0.0077&0.0009 & 0.016&0.003 & 
                 \multicolumn{2}{l}{0.0053}  & 0.019&0.06 \\
      urea     & 0.014&0.001   & 0.013&0.001 &
                 \multicolumn{4}{c}{$0.013 \pm 0.001$} \\
      ethylene glycol & 0.054&0.005 & 0.10&0.01 & 0.046&0.006 & 0.085&0.01 \\
      \hline
    \end{tabular}
    \label{tab:permeability}
  \end{center}
\end{table}

\begin{figure}
  \centering\includegraphics[scale=0.8]{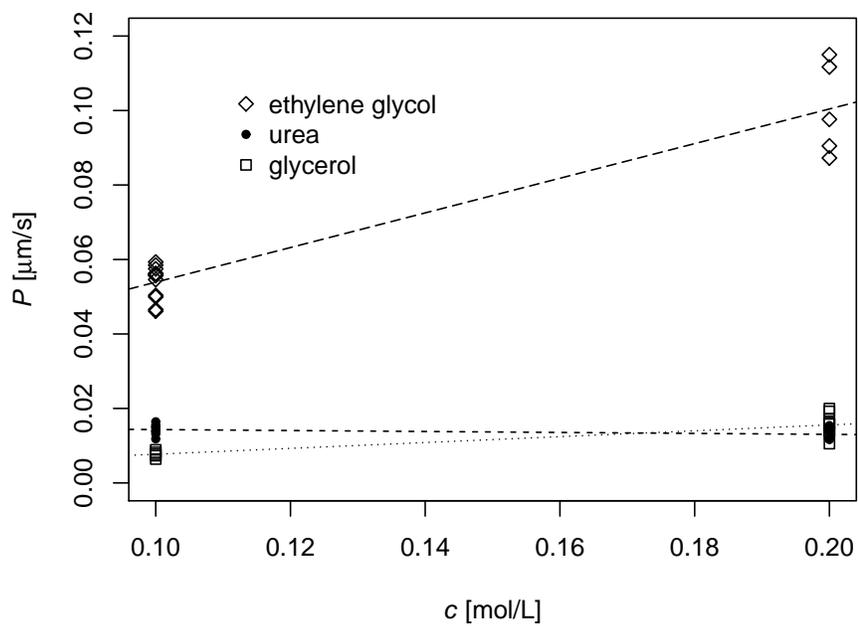}
  \caption{The apparent permeability of POPC membrane for glycerol,
    urea, and ethylene glycol, obtained by the ``burst train''
    analysis.  Fig.~6 in Ref.~\protect\cite{Peterlin:2012} shows
    analogous results obtained by the ``critical swelling'' analysis.}
  \label{fig:permeability-concentration}
\end{figure}

One can see that in the case of urea, the estimate for membrane
permeability is independent of solute concentration, which is
consistent with our finding in \cite{Peterlin:2012}.  On the other
hand, both polyols exhibit a concentration dependence of the
(apparent) membrane permeability, which is again in agreement with our
own recent findings \cite{Peterlin:2012}.  The trend is even more
apparent in Figure~\ref{fig:permeability-concentration}.  A possible
explanation for the concentration dependence of the apparent membrane
permeability is strong evidence for the affinity of polyols to
phospholipid headgroups \cite{Krasteva:2001,Pocivavsek:2011}, which
results in the solute concentration adjacent to the membrane being
larger than in the bulk solution, which consequently leads to an
apparent permeability coefficient which is larger than the actual one.
This affinity of polyols to phospholipid headgroups may also account
for a large scatter of published data on membrane permeabilities for
polyols.

\section{Discussion}

The existance of reproducible techniques for producing giant
unilamellar vesicles (GUVs) \cite{Angelova:1986,Pavlic:2011}, the
availability of phase-contrast microscopy and high-resolution digital
cameras, along with the efficient procedures for automated extraction
of vesicle contour from the images and determination of the radius of
vesicle cross-section \cite{Peterlin:2009a}, render the determination
of membrane permeability of GUVs a viable alternative to the
established techniques employing either dynamic light scattering on
sub-micrometer vesicles or planar lipid membranes.

GUVs were used before in studies of membrane permeability
\cite{Boroske:1981}.  The study by Boroske \emph{et al.}, however, was
designed to observe osmotic shrinking of GUVs, while in our study
vesicles swell osmotically (with the exception of the initial phase,
where a transient osmotic shrinking may be observed).  Osmotic
swelling of vesicles in a controlled manner guarantees a simpler
spherical geometry, through which both the vesicle area and its volume
are experimentally accessible variables.  By contrast, in the
experiment of Boroske \emph{et al.}, vesicle volume was the only
experimentally observable variable, and the authors had to come up
with a plausible explanation for the apparently ``missing'' membrane
area.

On a very superficial level, the continuous transition between the
``ironing'' and the stretching regime of a vesicle which swells
osmotically due to the permeation of a permeable solute might seem
similar to the ``minimal volume'' technique employed by Sha'afi
\emph{et al.} (\emph{cf.} Fig.~3 in \cite{Shaafi:1970} and Fig.~1 in
\cite{Peterlin:2012}).  However, one needs to be aware of the fact
that Fig.~3 in \cite{Shaafi:1970} shows vesicle volume as a function
of time, while Fig.~1 in \cite{Peterlin:2012} shows the radius of a
non-spherical vesicle cross-section, and throughout the course shown
in the diagram, vesicle volume is monotonously increasing.  In our
experiment, minimal vesicle volume appears at the top of the initial
bulge (visible around 40~s on the graph in
figure~\ref{fig:burst-seq}).  We have not attempted an analysis
analogous to the one performed by Sha'afi \emph{et al.}
\cite{Shaafi:1970} or Hill and Cohen \cite{Hill:1972}, because the
initial moment in which the vesicles are brought in contact with the
target solution is not defined as well as in the stopped flow
experiments.  The authors of earlier studies, on the other hand, seem
to be unaware of the critical transition between the ``ironing'' and
the stretching regime in the course of osmotical swelling of a
vesicle.

While offering a direct visualization of osmotic swelling, the method
described here has its limitations.  The three solutes tested here --
glycerol, urea, and ethylene glycol -- induce significant changes in
vesicle size on the timescale of minutes.  With solutes which are much
more or much less permeable, the rate of changes may be either too
fast for video camera, or so slow that other effects start to dominate
the experiment (\emph{e.g.,} convection in the measuring chamber).
Another limitation is imposed by the refractive index of the solution.
The refractive index of the target solution in general differs from
that of the vesicle interior, and the difference between the two
refractive indices increases with the increasing concentration of
solutions.  At high enough concentration, the halo around the vesicle
becomes blurred, and the algorithm used to determine vesicle contour
\cite{Peterlin:2009a} fails.  With our present equipment, we estimate
that concentrations significantly higher than 0.2~mol/L are outside
our reach.

Finally, we want to compare the results of two techniques dubbed
``burst train'' and ``critical swelling'' in Table
\ref{tab:permeability}.  As can be seen, they produce consistent
results which are within the experimental error of each other.  The
experimental error estimates for both methods are also approximately
the same.  The ``burst train'' technique is more forgiving towards
minor glitches, such as vesicle temporarily drifting slighly out of
focus, which results in an overestimate of vesicle radius (seen around
370~s in figure~\ref{fig:burst-seq}).  However, one needs to keep in
mind that the ``burst train'' technique requires approximately 5 times
as much data as the ``critical swelling'', and consequently, minor
glitches are more likely to occur during a longer period.  We
therefore lean towards recommending the ``critical swelling''
technique, which is less resource-greedy while giving the results with
the same precision.  Our recommended procedure is therefore to keep
recording the vesicle transfer up to the first burst (which is easy to
notice during the experiment), and carry out the analysis desribed in
Ref.~\cite{Peterlin:2012}.








\end{document}